\documentstyle[aps,pre,amssymb]{revtex}

\begin{document}

\twocolumn
\draft
\title{A Comment on the Tsallis Maximum Entropy Principle}
\author{Brian R. La Cour and William C. Schieve}
\address{Ilya Prigogine Center for Studies in Statistical Mechanics and Complex Systems \\
         Department of Physics \\
         The University of Texas at Austin \\
         Austin, Texas 78712}
\date{\today}
\maketitle

\begin{abstract}
Tsallis has suggested a nonextensive generalization of the
Boltzmann-Gibbs entropy, the maximization of which gives a generalized
canonical distribution under special constraints.  In this brief
report we show that the generalized canonical distribution so obtained
may differ from that predicted by the law of large numbers when empirical
samples are held to the same constraint.  This conclusion is based on
a result regarding the large deviation property of conditional
measures and is confirmed by numerical evidence.
\end{abstract}
\pacs{02.50.Cw, 05.20.Gg, 05.30.Ch}


\section{Introduction}

From considerations of multifractals, Tsallis \cite{Tsallis1988} was
led to conjecture a generalization of the Boltzmann-Gibbs entropy
given by
\begin{equation}
S_{q}(p) = \frac{1}{q-1}\left[1 - \sum_{i=1}^{m} p_i^{q} \right],
\label{eqn:Tsallis_entropy}
\end{equation}
where $p = (p_1,\ldots,p_m)$ is a probability distribution for a
discrete random variable with values $\epsilon_1,\ldots,\epsilon_m$
and $q$ is any real number different from one.  $S_{1}$ is defined to
be the usual Boltzmann-Gibbs entropy, in agreement with the limit $q
\rightarrow 1$.  (Boltzmann's constant is set to one.)  Non-Gibbsian
distributions are obtained by extremizing the Tsallis entropy under
special constraints, described below, while using $q$ as an adjustable
parameter.  The parameter $q$ typically has no direct physical
interpretation, but when it is used as an adjustable parameter the
resulting distributions can give surprisingly good agreement with
experimental data in a wide variety of fields \cite{Tsallis1999}.  In
a few cases, $q$ is uniquely determined by the constraints of the
problem and may thereby bear some physical interpretation
\cite{Beck2000,Lyra_and_Tsallis1998}.

Although the Tsallis entropy preserves all of the familiar
thermodynamic formalism, Curado \cite{Curado1999} has noted that this
is true of a much broader class of entropies.  Given the myriad of
possible entropy functions, one is led to ask why the Tsallis entropy
is special, and a natural place to look for answers is in the theory
of large deviations \cite{Ellis}, which gives a probabilistic
justification for the maximum entropy principle in terms of a unique
entropy function.  In this brief report we compare the probabilities
obtained by Tsallis's maximum entropy principle with the asymptotic
frequencies predicted by large deviation theory ({i.e.} the law of
large numbers) under similar constraints.  We find that the two do not
in general agree.


\section{Tsallis Maximum Entropy Principle}

If no constraints are imposed upon $p$ (other than that it be
nonnegative and normalized), $S_{q}$ is readily seen to be extremized
by $p_i = 1/m \equiv \mu_i$.  (The case $q = 0$ is special, as $S_{0}$
is a constant function.)  This conclusion, independent of $q$, agrees
with the usual Boltzmann-Gibbs result and corresponds to a
microcanonical ensemble.  If we view $\mu$ as a sampling distribution,
then the empirical distribution of frequencies obtained from a random
sample $x_1,\ldots,x_n$ converges to $\mu$ almost surely as $n$ grows
large.  This well-known result, originally due to Boltzmann
\cite{Ellis1999}, may be viewed as a example of the (strong) law of
large numbers.  Since $S_q$ has a global extremum at $\mu$, the
distribution predicted by extremizing $S_{q}$ agrees with the actual
asymptotic empirical distribution.

Placing additional constraints when extremizing $S_{q}$ may result in
a distribution dependent upon $q$, {i.e.} one at variance with that
predicted from the Boltzmann-Gibbs case $q = 1$.  As a generalization
of the internal energy constraint, Tsallis \cite{Tsallis1998} has
suggested the following constraint be used when extremizing $S_{q}$:
\begin{equation}
\sum_{i=1}^{m} (\epsilon_i - u)p_i^{q} = 0,
\label{eqn:q-expectation}
\end{equation}
where $u$ is a given fixed constant.  For $q = 1$ this of course
reduces to the usual expectation value constraint.  By extremizing
(\ref{eqn:Tsallis_entropy}) subject to (\ref{eqn:q-expectation}), one
obtains a solution in general different from the Boltzmann
distribution.  This solution is given explicitly by
\begin{equation}
p_i \propto \left[1-(1-q)\alpha(\epsilon_i-u)\right]^{1/(1-q)},
\label{eqn:q-canonical}
\end{equation}
where $\alpha$ is chosen such that Eqn.\ (\ref{eqn:q-expectation}) is
satisfied.  It has been noted that this explicit form of the
distribution appears to be more numerically robust than the more
common implicit form, for which $\alpha = \beta/\sum_{j=1}^{m}p_j^q$
\cite{Lima_and_Penna1999}.

For $q = 1$ the constraint on the expectation may be interpretation as
a constraint on the sample mean, the two being equivalent for large
samples.  Thus, if we consider random samples $x_1,\ldots,x_n$ from
$\mu$ which satisfy
\begin{equation}
\frac{1}{n} \sum_{k=1}^{n} x_k = u,
\end{equation}
then the empirical distributions of such samples will approach the
Boltzmann distribution $p_i \propto e^{-\alpha \epsilon_i}$ as $n$
grows large.

The question arises whether a similar interpretation may be made of
the constraint in Eqn.\ (\ref{eqn:q-expectation}) for $q \neq 1$ and,
more importantly, whether the resulting empirical distribution
converges to that given by Eqn.\ (\ref{eqn:q-canonical}).  As our
observable is discrete, let $f_{n,i}(x_1,\ldots,x_n)$ denote the
observed frequency of $\epsilon_i$ in the sample $x_1,\ldots,x_n$.
(There is no obvious interpretation for continuous values.)  We may
interpret Eqn.\ (\ref{eqn:q-expectation}) to mean
\begin{equation}
\sum_{i=1}^{m} (\epsilon_i-u) f_{n,i}(x_1,\ldots,x_n)^q = 0.
\label{eqn:empirical_constraint}
\end{equation}
We will show that random samples drawn from $\mu$ which satisfy Eqn.\
(\ref{eqn:empirical_constraint}) do not in general give rise to
empirical distributions which converge to the Tsallis prediction of
Eqn.\ (\ref{eqn:q-canonical}).


\section{Conditional Convergence of the Empirical Distribution}

The general problem we are considering is the convergence in
probability of the empirical frequencies $f_{n} =
(f_{n,1},\ldots,f_{n,m})$, where $f_{n}$ is a random vector with
domain $\{\epsilon_1,\ldots,\epsilon_m\}^n$ taking values in the
convex set ${\mathcal{P}} = \{p \in {\mathbb{R}}^m: p_i \ge 0,
\sum_{i=1}^{m} p_i = 1\}$.  Unconstrained, an infinite random sample
$x_1,x_2,\ldots,$ from $\mu$ gives rise to a sequence of empirical
frequencies which converge in probability to $\mu$.  Sanov's theorem
\cite{Ellis1999} gives the large deviation rate function for this
convergence to be just the negative of the Boltzmann-Gibbs entropy:
\begin{equation}
I_{\mu}(p) = -S_{1}(p) - \log m.
\end{equation}
Loosely speaking, Sanov's theorem states that for $A \subseteq
{\mathcal P}$, $\mu^n[f_{n} \in A] \sim \exp[-n\inf_{p\in A}
I_{\mu}(p)]$ for large $n$ (cf. the Boltzmann-Einstein formula $W =
e^{S}$).  The asymptotic measure, $\mu$, is the unique minimum of the rate
function $I_{\mu}$, which is continuous and strictly convex.

When we impose additional constraints on $f_{n}$, the asymptotic value
changes from $\mu$ to a new distribution which minimizes $I_{\mu}$
under the added restrictions \cite{Ellis1999,LaCour_and_Schieve2000}.
If we condition on the sample mean for example, {i.e.}
\begin{equation}
\sum_{i=1}^{m} \epsilon_i f_{n,i}(x_1,\ldots,x_n) = u,
\label{eqn:fixed_mean}
\end{equation}
the resulting asymptotic distribution is no longer $\mu$ but the
canonical distribution $P_i \propto e^{-\beta \epsilon_i}$, where
$\beta$ satisfies
\begin{equation}
\sum_{i=1}^{m} \epsilon_i P_i = u.
\label{eqn:fixed_expectation}
\end{equation}
It is in this sense that finding the asymptotic empirical distribution
under (\ref{eqn:fixed_mean}) is equivalent to maximizing $S_{1}$ under
(\ref{eqn:fixed_expectation}).

More generally, imposing condition (\ref{eqn:empirical_constraint})
results in an asymptotic distribution which minimizes $I_{\mu}$
(maximizes $S_{1}$) subject to (\ref{eqn:q-expectation}).  This
distribution is given implicitly by
\begin{equation}
P_i \propto \exp\left[-\beta(\epsilon_i-u)P_i^{q-1}\right],
\label{eqn:asymptotic_distribution}
\end{equation}
where $\beta$ is such that Eqn.\ (\ref{eqn:q-expectation}) is
satisfied with $p$ replaced by $P$.  Comparison with Eqn.\
(\ref{eqn:q-canonical}) shows that both $p$ and $P$ will agree when $q
\rightarrow 1$.


\section{Comparison of the Two Distributions}

For $q = 0$, Eqn.\ (\ref{eqn:q-canonical}) gives $p_i =
[1-\alpha(\epsilon_i-u)]/m$, with $\alpha$ unrestricted, while Eqn.\
(\ref{eqn:asymptotic_distribution}) implies $P_i = 1/m$.  Clearly both
agree if $\alpha$ is arbitrarily chosen to be zero.  However, as we
have noted $S_{0}$ is a constant function, so the entropy
extremization procedure may be expected to break down in this case.

Taking $u$ to be the equilibrium value $u_* = \sum_{i=1}^{m}
\epsilon_i/m$ also results in general agreement between $p$ and $P$
for all $q \neq 0$.  Indeed, by choosing $\alpha = \beta = 0$ we see
that $p_i = 1/m$ is the unique solution for both Eqn.\
(\ref{eqn:q-canonical}) and Eqn.\ (\ref{eqn:asymptotic_distribution}).
This agreement simply reflects that fact that both $S_1$ and $S_{q}$
have the same global extremum.

When $m=2$ the two constraints are sufficient to uniquely determine
the distribution, and for this reason general agreement is also
expected.  In particular we find
\begin{equation}
p = P \propto \left((\epsilon_2-u)^{1/q}, (u-\epsilon_1)^{1/q}\right),
\end{equation}
assuming $\epsilon_1 < \epsilon_2$ and $q \neq 0$.  It is readily
verified that Eqn.\ (\ref{eqn:q-expectation}) is satified.  By solving
for $\alpha$ and $\beta$, Eqns.\ (\ref{eqn:q-canonical}) and
(\ref{eqn:asymptotic_distribution}), respectively, may be satisfied as
well.

Disagreement between $p$ and $P$ is therefore expected when $m \ge 3$.
To show this explicitly, we may compute $p$ from Eqn.\
(\ref{eqn:q-canonical}) for an arbitary $u$ and then search for a
value of $\beta$ such that Eqn.\ (\ref{eqn:asymptotic_distribution})
is satisfied when $p$ is substituted for $P$.  The claim is that a
single $\beta$ cannot always be found which satisfies this equation
for all values of $i$ when $m \ge 3$.

The case $q=1/2$ is particularly amenable to analytic study
\cite{Rebolla-Neira1998} and appears in an early application of the
Tsallis entropy to turbulence in a two-dimensional electron plasma
\cite{Boghosian1995}.  For this case, Eqn.\ (\ref{eqn:q-canonical})
may be solved explicitly in terms of $u$ to obtain
\begin{equation}
p_i \propto \left[\frac{1}{m}\sum_{j=1}^{m}(\epsilon_j-u)^2 - (\epsilon_i-u)(u_*-u)\right]^2.
\end{equation}
Using a given value of $u$ and the corresponding $p$ given above, we
then consider zeros of the functions $d_i$, where
\begin{equation}
d_i(\beta) = \frac{\exp\left[-\beta(\epsilon_i-u)p_i^{-1/2}\right]}{\sum_{j=1}^{m} \exp\left[-\beta(\epsilon_j-u)p_j^{-1/2}\right]} - p_i,
\end{equation}
for $i=1,\ldots,m$.  A plot of these functions is shown in Fig.\
\ref{fig:zeros} for selected parameter values.  The failure of all
three graphs to have a zero at the same value of $\beta$ indicates
that $p$ and $P$ are in this case distinct.

From this example one can derive a general necessary condition for
agreement with $P$.  Suppose that for given $q$, $\epsilon$, and $u$
there exists a simultaneous solution to both Eqns.\
(\ref{eqn:q-canonical}) and (\ref{eqn:asymptotic_distribution}).
(More generally, $p$ may be any probability distribution satisfying
Eqn.\ (\ref{eqn:q-expectation}).)\ Substituting the former into the
latter we find
\begin{equation}
p_i = \exp[-\beta(\epsilon_i-u)p_i^{q-1}] / Z(\beta),
\label{eqn:impossible}
\end{equation}
where
\begin{equation}
Z(\beta) = \sum_{j=1}^{m}\exp[-\beta(\epsilon_j-u)p_i^{q-1}].
\end{equation}
The value of each $p_i$ is fixed in terms of the given parameters, so
a single value of $\beta$ must simultaneously satisfy Eqn.\
(\ref{eqn:impossible}) for $i=1,\ldots,m$.  If any $p_i = 0$ then
Eqn.\ (\ref{eqn:impossible}) cannot possibly be satisfied, so suppose
all $p_i$ are nonzero.  For any given $j \neq i$,
\begin{equation}
\beta = -[\log p_j + \log Z(\beta)] / [(\epsilon_j-u)p_j^{q-1}].
\end{equation}
Substituting this expression back into Eqn.\ (\ref{eqn:impossible})
gives
\begin{equation}
\log Z(\beta) = \frac{(\epsilon_i-u)p_i^{q-1} \log p_j - (\epsilon_j-u)p_j^{q-1} \log p_i}{(\epsilon_j-u)p_j^{q-1} - (\epsilon_i-u)p_i^{q-1}}.
\label{eqn:impossible_too}
\end{equation}

The RHS of Eqn.\ (\ref{eqn:impossible_too}) is invariant under the
interchange of $i$ and $j$, so it has at most $m(m-1)/2$ distinct
values.  The LHS, of course, is the same for all choices of $i$ and
$j$.  Now, the RHS will be independent of the choice of $i$ and $j$ if
either (1) $q=1$, (2) $m=2$, or (3) $p_i=p_j$ for all $i$ and $j$, the
latter being equivalent to $u=u_*$, which is equivalent to $\alpha=0$.
Assuming none of these three conditions hold, the RHS must be the same
for all choices of $i$ and $j$ if indeed $p=P$.  This gives a
necessary condition for agreement.


\section{Discussion}

We have compared the probability distribution over $m$ states
predicted from Tsallis's maximum entropy principle, which constrains
the normalized $q$-expectation to a value $u$, to the asymptotic
frequencies when the empirical $q$-expectation is similarly
constrained.  The two will always agree if either (1) $q=1$, (2)
$m=2$, or (3) $u=u_*$.  A specific example for which $q=1/2$ and $m=3$
was used to demonstrate numerically that the two distributions may be
different.  For the case in which none of these three conditions hold,
we derived a necessary condition to be satisfied by any candidate
distribution in order that it be identical to true asymptotic
distribution.

From the point of view of large deviation theory, the maximum entropy
principle specifies the overwhelmingly most probable distribution to
be realized by a large-sample empirical distribution under given
constraints.  The uniqueness of the rate function in large deviation
theory implies that the Boltzmann-Gibbs entropy plays a special role
in determining this most likely distribution.  For this reason, novel
entropy functions such as that proposed by Tsallis may give results
which are at variance with actual sample frequencies except, as
observed, in some special cases.




\begin{figure}
\caption{Plot of $d_i(\beta) = \rho_i(\beta) - p_i$ for $\epsilon =
(0,1,2)$, $q=1/2$, and $u=7/11$, for which $p = (289, 121, 25)/435$.
The positive roots are found numerically to be $0.514509$, $0.637715$,
$0.360903$ for $i=1,2,3$ respectively.}
\label{fig:zeros}
\end{figure}


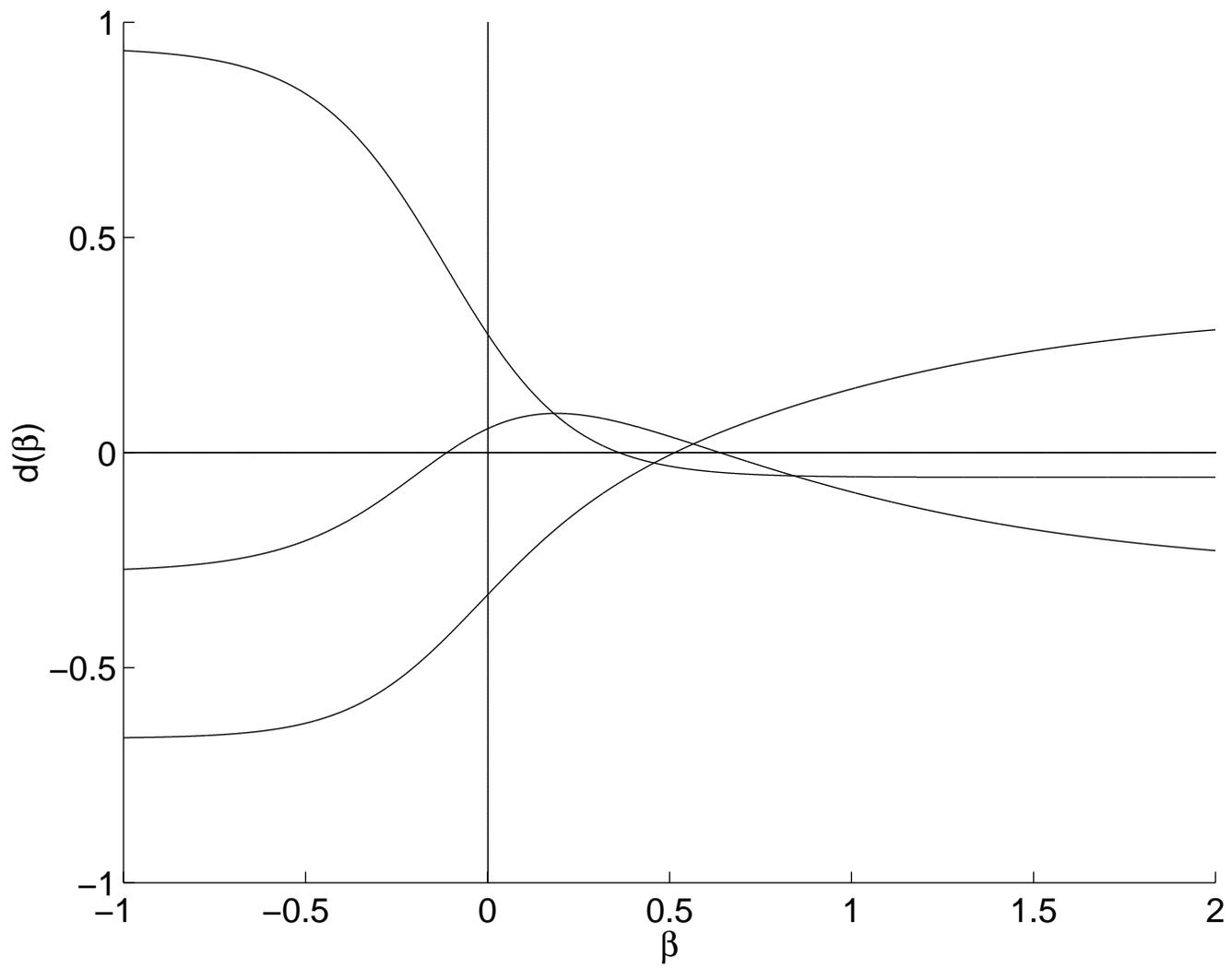
\end{document}